\documentclass[11pt,a4paper]{article}
\usepackage{moreverb}
\usepackage{float}

\usepackage[dvipdfmx]{graphicx}  
\usepackage{ascmac}
\usepackage{amsmath}
\usepackage{subcaption}
 
\usepackage{array}   

\newcommand{\fulltoday}{\number\day\space \ifcase\month\or
January\or February\or March\or April\or May\or June\or
July\or August\or September\or October\or November\or December\fi
\space\number\year}

\begin{document}
\title{Stock Market  Market Crash of 2008: an empirical study of the deviation of share prices from company fundamentals}

\author{Taisei Kaizoji and Michiko Miyano \\
Graduate School of Arts and Sciences,\\ 
International Christian University\footnote{Graduate School of Arts Sciences, International Christian University, 3-10-2 Osawa, Mitaka, Tokyo, 181-8585, Japan. Corresponding autor: Taisei Kaizoji kaizoji@icu.ac.jp} } 
\maketitle
\begin{center}
Abstract 
\end{center}
The aim of this study is to investigate quantitatively whether share prices deviated from company fundamentals in the stock market crash of 2008. For this purpose, we use a large database containing the balance sheets and share prices of 7,796 worldwide companies for the period 2004 through 2013. We develop a panel regression model using three financial indicators--dividends per share, cash flow per share, and book value per share--as explanatory variables for share price. We then estimate individual company fundamentals for each year by removing the time fixed effects from the two-way fixed effects model, which we identified as the best of the panel regression models. One merit of our model is that we are able to extract unobservable factors of company fundamentals by using the individual fixed effects. 

   Based on these results, we analyze the market anomaly quantitatively using the divergence rate--the rate of the deviation of share price from a company's fundamentals. We find that share prices on average were overvalued in the period from 2005 to 2007, and were undervalued significantly in 2008, when the global financial crisis occurred.  Share prices were equivalent to the fundamentals on average in the subsequent period. Our empirical results clearly demonstrate that the worldwide stock market fluctuated excessively in the time period before and just after the global financial crisis of 2008. 
	
\section{Introduction} 
The stock market crash of 2008 is one of the largest stock market crashes in the history of capitalist economies. During the period from 2004 through 2013, the Dow Jones Industrial Average hit a high of 14,164.43 on October 9, 2007. For the eight trading days between October 1 and October 10, 2008, the DJIA fell continuously from 10,831.07 to 8,451.19, a 22.11 percent decline. The DJA hit a bottom of 6,594.44 on March 5, 2009. In less than 18 months, the index had declined more than 50 percent. The crisis was not limited to the US market. Markets worldwide were simultaneously in free fall. Figure 1 shows the mean of the logarithmic share price for 7,796 worldwide companies in the period 2004 through 2013. This value hit 1.71 in 2007, and in 2008 hit a bottom of 1.26. The mean share price declined by 36 percent in one year.  

The liquidity crunch in U.S. and European short-term money markets began with an incident in August 2007 involving the “complete evaporation of liquidity” of three hedge funds invested in U.S. asset-backed securities (ABS) affiliated with BNP Paribas, one of the largest banks in France. U.S. and European financial institutions that provided liquidity support for the redemption of asset backed commercial paper (ABCP) were obliged to raise funds. Consequently, liquidity pressures in funding markets rose, spurring a liquidity crisis in short-term money markets.  Amid this abrupt tightening of global financial markets, which triggered Lehman Brothers, a large investment bank, to file for bankruptcy in September 2008, investors in large numbers rapidly withdrew their funds from the stock markets, causing severe global disruptions. 

In an efficient market (Fama 1970), stock price volatility is linked to changes in company fundamentals. To explore this notion, numerous attempts have been made by scholars to determine whether stock price volatility systematically exceeds levels which could be justified by changes in fundamentals. (See Shiller 1981, LeRoy and Porter 1981, Mehra and Prescott 1985, De Bondt and Thaler 1985, Fama and French 1992, Jegadeesh and Titman 1993, etc.) The issue remains controversial. 

This paper examines the question of whether the stock market crash of 2008 was an efficient response to financial shocks that was in line with fundamentals or was caused by investor panic. In order to produce estimates of the fundamentals required for our study, we construct a panel regression model using three financial indicators — dividends per share, cash flow per share, and book value per share— as explanatory variables for share price. These financial indicators are the representative variables commonly used to evaluate a firm's business performance. We perform the panel analysis using a large database gleaned from the balance sheets of 7,796 of the world's largest listed companies over a 10-year period (2004-2013). The two-way fixed effects model was selected as the best panel regression model for our work, based on standard tests for panel regression models. 

The two-way fixed effects model has two fixed effects: the individual fixed effects that account for an individual company's heterogeneity, including such factors as the company's diversity of corporate governance and the quality of its employees; and the time fixed effects that indicate variables that fluctuate over time but are fixed across companies. The time fixed effects reflect various shocks, including financial shocks. 

We define fundamentals as the theoretical value that omits the time fixed effects from our estimated regression model. One advantage of our model is that it can capture unobservable factors explaining company fundamentals. We investigated the distributions of the divergence rate, which is defined as the logarithmic difference between the share price and the fundamentals, and found that share prices deviated substantially from company fundamentals in the period 2006 to 2008. The distributions of the divergence rate deviated in the positive direction in the boom period from 2006 through 2007, but shifted significantly from the positive side to the negative side in 2008. It is clear that share prices (on average) were overvalued against the fundamentals during the boom period from 2006 to 2007, while in 2008 they were significantly below the fundamentals. In addition, the distributions of the divergence rate were negatively skewed and leptokurtic as compared to the distributions in other periods. It is notable that the negative skewness and leptokurtosis of the distributions of the divergence rate is indicative of the danger of a bubble. We conclude that the bubble of 2006 and 2007, and the subsequent crash of 2008, cannot be linked to changes in company fundamentals, but rather was likely caused by factors such as the psychological panic of investors. 

This paper is organized as follows: Section 2 describes the data used in this study; Section 3 discusses the panel data regression model for company fundamentals; Section 4 examines the divergence rate, that is, the deviation of share prices from the fundamentals; Section 5 gives concluding remarks.

\section{Data} 
The data for this paper were collected from the OSIRIS database provided by Bureau van Dijk containing company financial statements and reports for nearly 80,000 companies listed around the world, 

One difficulty with international comparisons of financial statements collected from multiple sources is that the formats of the statements tend to be different by source and country. As a result, inconsistencies across sources can cause problems of comparability of the data. Mismeasurements can lead to biased results. The merit of OSIRIS is that it provides information on company financials in a standardized format expressed in local currencies and US dollars. Thus, one can use OSIRIS financial data for worldwide companies without encountering comparability problems. 

   In this paper, we perform a statistical investigation of stock prices and financial indicators per share for 7,796  companies over the 10-year period from 2004 through 2013. As is evident, the database contains data for time periods before and after the global financial crisis. 

\section{Panel data analysis}
In this section, we examine the panel data described in the previous section. The aim of the panel data analysis is to build a model estimating company fundamentals, which establish the fair value of a share estimated from a company's balance sheet. 

  To calculate company fundamentals, we use three financial indicators as the explanatory variables for share price—{\itshape dividends per share, cash flow per share, and book value per share}. These three indicators are commonly used by financial professionals and investors in fundamental analysis as a tool for identifying the divergence of share price in the market from the intrinsic value of a company. These same three financial indicators were used as the explanatory valuables for share price in our previous study (kaizoji and Miyano 2016). In that study, a simple cross-sectional analysis of share prices per year was conducted to explain the power law for share price. A weakness of such a cross-sectional analysis is that it is unable to consider company-specific characteristics. Given the presence of unobserved heterogeneity among companies—qualitative factors such as corporate governance and the quality of company employees—we develop the panel regression model in this study in order to express this unobserved heterogeneity. 
  
  Based on our analysis of the panel data, a two-way fixed effects model of share prices was selected as the best share price model. The two-way fixed effects model controls for (i) individual fixed effects which control for unobserved factors that differ between companies but are constant over the 10-year period for each company, and (ii) period fixed effects which control for unobserved factors that are shared by all companies at a specific point in the year and are not accounted for by the three financial indicators. 

\subsection{The explanatory variables of share price}
Dividends per share, cash flow per share, and book value per share are the financial indicators commonly used to evaluate fundamentals. Most elementary stock valuation methods are based on company profits and shareholder equity. We introduce briefly these financial indicators. 

  In our OSIRIS database, cash flow per share is defined as net income plus depreciation divided by the number of outstanding shares of company stock. Earnings per share (EPS) is often used in fundamental analysis as an alternative to cash flow per share. However, we prefer cash flow per share to earnings per share since, as many analysts point out, ``earnings can be manipulated more easily than cash flow.'' Dividends per share is calculated as the total dividends received by shareholders for each outstanding share of the company. Book value per share is the amount of money that a shareholder would receive if a company were to liquidate. Book value per share is often used as a measure to judge whether a share is overvalued or undervalued. If the share price exceeds book value per share, then the share price may be overvalued in the stock market, and vice versa.

\subsection{Panel data regression models}
We performed our panel data analysis using the financial indicators introduced in the previous section. All of the distributions of share price, dividends per share, cash flow per share, and book value per share are highly skewed. Therefore, we used a logarithmic transformation of the variables. The log transformation can be useful in satisfying the regression assumptions for such panel data. The panel data regression model is written as
\begin{equation}
lnY_{it}=a+b_{1}lnX_{1,it}+b_{2}lnX_{2,it}+b_{3}lnX_{3,it}+u_{it}  \quad i=1,\dots, N; \quad t=1,\dots T
\end{equation}    
 where $Y_{it}$  denotes the dependent variable (the share price) for company $i$   in year $t$  ; 
$a$  denotes a constant; $X_{1,it}$ is the dividends per share of company  $i$ in year $t$ ; $X_{2,it}$  is the cash flow per share of company $i$   in year $t$ ; $X_{3,it}$  is the book value per share of company $i$   in year $t$  ; $u_{it}$  denotes the error term. 
We estimate the model in equation (1) using the Panel Least Squares method. In the panel regression mode, the error term, $u_{it}$  can be assumed to be divided into a pure disturbance term and an error term due to other factors. Assuming a two-way error component model with respect to error, the factors other than disturbance are (i) factors due to unobservable individual effects, and (ii) factors due to unobservable time effects. That is, the error term can be written as
 \begin{equation}
u_{it}=\mu_{i}+\gamma_{t}+\epsilon_{it} 
 \end{equation}
                                                                                                          
where $\mu_{i}$   denotes unobservable individual effects, $\gamma_{t}$  denotes unobservable time effects, and $\epsilon_{it}$  denotes pure disturbance. 

   If both $\mu_{i}$   and $\gamma_{t}$  are equal to zero, equation (1) is estimated using the pooled OLS method. If either $\mu_{i}$  or $\gamma_{t}$  is equal to zero, equation (2) is a one-way error component model. If both $\mu_{i}$   and $\gamma_{t}$  are not equal to zero, equation (2) is a two-way error component model. There are two estimation methods for estimating the error term in equation (2). One is fixed effects estimation and the other is random effects estimation. Therefore, the available estimation models are a pooled OLS, an individual fixed effects model, a time effects model, a two-way fixed effects model, an individual random effects model, a time random effects model, and a two-way random effects model.\footnote{Tow-way random effects model is unavailable since we use unbalanced panel data.}
   
   We estimated the models described above and found that the two-way fixed effects model was selected as the best model after appropriate model selection tests. The model selection tests used in this study include the likelihood ratio test and F-test for the selection of the pooled OLS model vs the fixed effects model, and the Hausman test for the selection of the random effects model vs the fixed effects model. The selection test for the pooled OLS model vs the random effects model is based on the simple test proposed by Wooldridge (2010).\footnote{Wooldridge(2010,p299) proposed the method that uses residuals from the pooled OLS to check the existence of serial correlation.} 
 
   The two-way fixed effects mode is written as
 \begin{gather}
lnY_{it}=a+b_{1}lnX_{1,it}+b_{2}lnX_{2,it}+b_{3}lnX_{3,it}+\epsilon_{it}   \notag \\
a=a_{0}+\mu_{i}+\gamma_{t} 
\end{gather}  
where $a_{0}$  is a constant term common to all companies, $\mu_{i}$    denotes the individual fixed effects, and $\gamma_{t}$  denotes the time fixed effects, $\mu_{i}$  is constant toward time series and $\gamma_{t}$   is constant toward cross section.  $\epsilon_{it}$  is the pure disturbance. The individual fixed effects,  $\mu_{i}$  , accounts for an individual company's heterogeneity and includes such factors as the company's diversity of corporate governance and the quality of its employees.  The time fixed effects, $\gamma_{t}$  , indicates variables that fluctuate over time but are fixed across companies. The time fixed effects reflect various shocks, including financial shocks. 

   Table 1 presents the estimates produced for the two-way fixed effects model. The first line shows the estimated intercept and estimates of the coefficients of the explanatory variables—dividends per share, cash flow per share, and book value per share. The second line shows the standard error of the estimates modified using the White period method, since we detected heteroscedasticity for the residuals and serial correlation of the residuals in the two-way fixed effects model. 
   
\begin{table}
\begin{center}
\caption{ Results of estimates for the two-way fixed effects model .}
\begin{tabular}{lcccc} \ 
               & $a_{0}$ & $b_{1}$ & $b_{2}$ & $b_{3}$ \\ \hline
coefficient   &1.485 & 0.137   & 0.208   &0.378   \\
 Std. error    &0.032 & 0.007   & 0.007   & 0.019  \\
 t-Statistic   & 46.07 & 19.45 & 28.46    & 19.55  \\
 $p$-value  & 0.000 & 0.000 & 0.000    & 0.000  \\  \hline
 R-squared             & 0.969 &       &             &      \\
 $p$-value (F-statistic) & 0.000 &        &    &   \\  \hline
\end{tabular}
\end{center}
\end{table}

The coefficients of the three financial indicators have a positive sign and are statistically significant. The $p$-values for all explanatory variables are near zero, indicating that the null hypothesis—that the coefficient is equal to zero—can be rejected in each case. The $p$-value for the overall F-test is also very close to zero and the R-squared value (0.97) is high. The R-squared statistic has a high value for all of the regression models, indicating that the regression model explains the variation in stock prices very well. More concretely, it means that the theoretical value explains 97 percent of the total variation in the stock prices about the average. The positive sign of the estimates justifies the fundamental analysis. For example, the coefficient of the dividend per share suggests that a 1 percent increase in dividend per share can be associated with a 0.13 percent rise in the share price. 

   Estimates of the two-way fixed effects model for share price, $ln\hat{Y}_{it}$  are written as 
\begin{equation}
ln\hat{Y}_{it}=\hat{a}+\hat{b}_{1}lnX_{1,it}+\hat{b}_{2}lnX_{2,it}+\hat{b}_{3}lnX_{3,it} 
\end{equation} 
We call $\hat{Y}$  the theoretical value of share price. Figure 2 is a scatter diagram of the theoretical value of the logarthmic share price plotted against the actual logarthmic share price. Figure 2 suggests that the relationship between theoretical value and actual share price is highly positive.

Figure 3 shows that the relative frequency distribution of the individual fixed effects (which are constant over time) for 6,209 companies.\footnote{Since 7,796 companies used in this study are unbalanced panel data for regression model, we obtained individual fixed effects of 6,209 companies.} The mean of the individual fixed effects is -0.054; the standard error is 0.01. The distribution indicates a wide heterogeneity in the unobservable capability of the studied companies.

Figure 4 shows the time effects reported separately for each year. The movement of these time fixed effects is considered to be the result of temporal shocks to the stock market. We discuss the time fixed effects further in Section 4.

\subsection{Company fundamentals}
To estimate the fundamentals of individual companies, we eliminate the time fixed effects from the two-way fixed effects model, while retaining the individual fixed effects. The reason for eliminating the time fixed effects term is that these effects are considered to be the effects of temporal financial and economic shocks on share price. We retained the individual fixed effects because these effects represent the individual company's unobserved heterogeneity as reflected in its share price. Therefore, we define the logarithmic form of a company's fundamentals as 
\begin{equation}
ln\tilde{Y}_{it}=\hat{a_{0}}+\hat{\mu_{i}}+\hat{b}_{1}lnX_{1,it}+\hat{b}_{2}lnX_{2,it}+\hat{b}_{3}lnX_{3,it} 
\end{equation}

where $\tilde{Y}_{it}$  denotes the fundamentals of company $i$  in year $t$. 

   This model of company fundamentals serves to further our purpose of investigating the deviation of a company's share price from its fundamentals. The model is different from other fundamental analysis and offers substantial value. The estimates of the coefficients in equation (5), $\hat{a_{0}}$  ,$\hat{\mu}_{i}$ ,$\hat{b}_{1}$  ,$\hat{b}_{2}$ ,and $\hat{b}_{3}$ , are constant over time. Therefore, if we can obtain values for dividends per share, cash flow per share, and book value per share for a company, we can easily estimate the company's fundamentals. 

\subsection{Divergence rate of share price from company fundamentals}
We use the company fundamentals ($\tilde{Y}$) model for the 10-year period from 2004 through 2013 to pursue our primary goal: to investigate the deviation of share price from company fundamentals. The divergence rate between share price and company fundamentals is defined as 
\begin{equation}
D_{it}=lnY_{it}-ln\tilde{Y}_{it}
\end{equation}
where $Y_{it}$   denotes the share price of company $i$  in year $t$, and $\tilde{Y}_{it}$ denotes the fundamentals of company $i$   in year $t$. 

The divergence rate, $D_{it}$ , for company $i$'s share prices is the rate of change between company $i$'s share price and the company $i$'s fundamentals in year $t$. We calculate the divergence rate $D_{it}$  for each company for each year. Table 2 shows basic statistics for the divergence rates for each year over the 10-year period from 2004 through 2013. Figure 5 shows the mean of the divergence rates and indicates that there is substantial variation in the divergence rate over time. 
   
\begin{table}
\begin{center}
\caption{Basic statistic of divergence rate for each year.}
\begin{tabular}{lccccc} \
 Year & Mean & Std. Dev. & Kurtosis & Skewness & Observations \\  \hline
2004 & 0.063 &0.368 & 4.3 & 0.0 &4807   \\
2005 & 0.148 &0.390 & 3.7 &-0.3 &4884   \\
2006 & 0.161 &0.339 &60.7 &-3.9 &4822   \\
2007 & 0.109 &0.386 &61.8 &-4.0 &4914   \\
2008 &-0.342 &0.356 & 3.6 &-0.3 &4383   \\
2009 &-0.048 &0.275 & 4.0 & 0.6 &4364   \\
2010 &-0.007 &0.276 & 1.9 & 0.6 &4675   \\
2011 &-0.103 &0.265 & 2.7 &  0.0&4719  \\
2012 &-0.055 &0.302 &16.3 &-0.8 &4770   \\
2013 & 0.045 & 0.330& 2.1  & 0.3& 4823    \\ \hline
\end{tabular}
\end{center}
\end{table}

As can be seen here, the mean divergence rate is more than 0.1 for the years 2005 to 2007, during which time the world economy and financial markets enjoyed a boom period. The mean then fell sharply, from 0.1 in 2007 to minus 0.34 in 2008, amid the global financial crisis. The implication is that stocks were, on average, bought excessively from 2005 through 2007, and, on average, overly sold in 2008 relative to the fundamentals. It is clear that the reason for this precipitous fall in the average divergence rate in 2008 was the global financial crisis of that year.

      We also investigate the distribution of the divergence rate. Figure 6 shows the distribution for the period from 2006 to 2008, which includes the period before and during the global financial crisis. The figure shows clearly that the distribution of the divergence rate shifted drastically towards the minus side from 2007 to 2008. On the other hand, Figure 7 shows the distribution of the divergence rate for the years 2009 through 2013, which is not a period of bubble and panic but rather is relatively normal. During the stock market boom (2006-2007), the distributions of the divergence rate are pulled in the positive direction, then suddenly move in the opposite (negative) direction in 2008. In other words, the distribution of the divergence rate is unimodal in normal times, shifting to bimodal in times of bubble and crash. Some authors draw an analogy between non-equilibrium phase transitions and the collapse of a speculative bubble. (See Chowdhury and Stauffer 1999, Kaizoji 2000, Boland 2009). On the assumption that trading in the stock market can be described by an Ising spin model, a phase transition from a bull market to a bear market can be characterized as a stock market crash. Our empirical finding is in agreement with this theoretical hypothesis.  (See Kaizoji 2000)

\section{Conclusion}
In this paper, we examine the deviation of share price from company fundamentals. Using company balance sheet data, we propose a panel regression model of share prices for 7,796 companies listed worldwide over the 10-year period from 2004 to 2013. We find that a two-way fixed effects model of share price that uses three financial indicators—dividends per share, cash flow per share, and book value per share—as the explanatory variables fits very well to the panel share price data. We estimate company fundamentals by removing the time fixed effects from the two-way fixed effects model, recognizing that the time fixed effects represent the effect of temporary shocks on share price. One advantage of our model of fundamentals is that we are able to quantitatively estimate unobservable factors in company fundamentals using the individual fixed effects. More concretely, by using our model (i) the parameters of the model of fundamentals can be estimated with a panel data regression, and (ii) unobserved heterogeneity among companies can be quantified by using the individual fixed effects. Our model of fundamentals is of value both to researchers and to investors. 

Having established an effective model for determining company fundamentals, we then investigate the divergence rate—measuring the deviation of a company's share price from the company's fundamentals. The mean divergence rates are positive in the years from 2005 to 2007 but declined drastically to a large negative value in 2008. These results suggest that share prices were overvalued, on average, during the period of the financial boom, but were significantly undervalued, on average, due to the stock market crash caused by the global financial crisis in 2008. The results of our empirical study provide evidence of excessive volatility. 
   
The financial crisis of 2008 reduced the world economy to a deep and prolonged recession and raises the question of how financial diseases are disseminated. Examining the mechanisms for such propagation is of interest and will provide an added focus for our future research. 
 future. 

\section{Acknowledgment} 
This research was supported by JSPS KAKENHI Grant Number 2538404, 2628089.

\break


\begin{figure}
\begin{center}
\includegraphics[width=8cm]{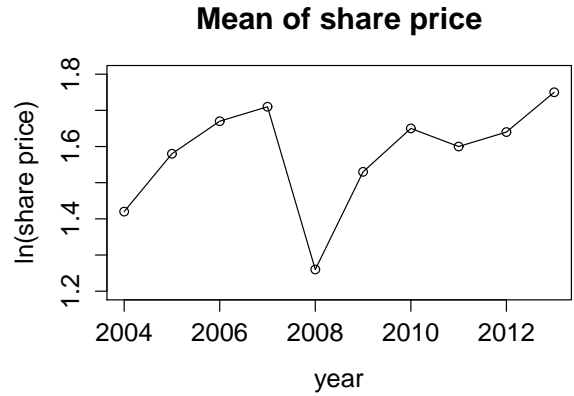}
\caption{Mean of logarthmic share price: The  shows the time series of the mean of logarithm of share price  of 7,796 world-wide companies in the period of 2004 through 2013. }
\end{center}
\end{figure}

\begin{figure}
\begin{center}
\includegraphics[width=8cm]{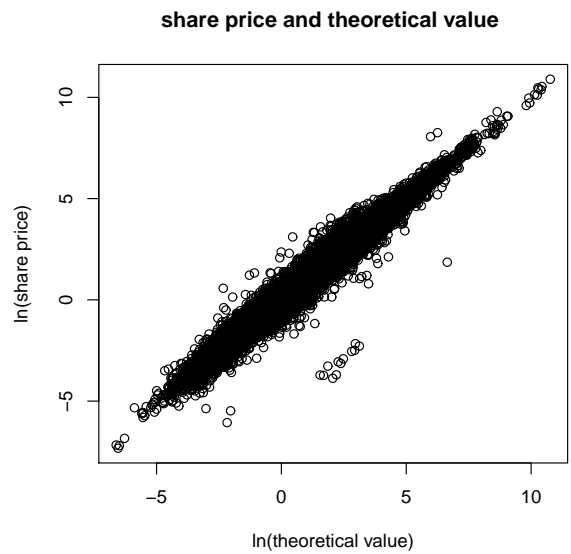}
\caption{Scatter diagram of the theoretical value of the share price against the actual share price}
\end{center}
\end{figure}
\begin{figure}
\begin{center}
\includegraphics[width=8cm]{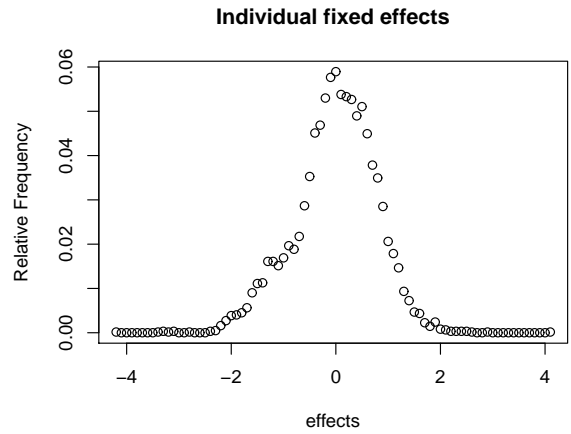}
\caption{The relative frequency distribution of the individualfixed effects , $\mu_{i}$ in the regression model (4)}
\end{center}
\end{figure}

\begin{figure}
\begin{center}
\includegraphics[width=8cm]{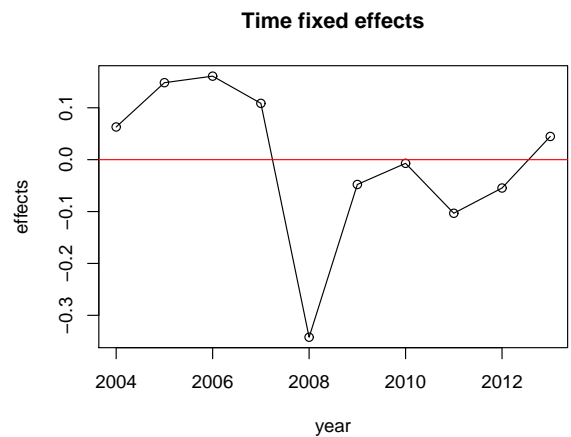}
\caption{The time series of the time fixed effects The relative frequency distribution of the individualfixed effects , $\gamma_{t}$ in the regression model (4)}
\end{center}
\end{figure}
\begin{figure}
\begin{center}
\includegraphics[width=8cm]{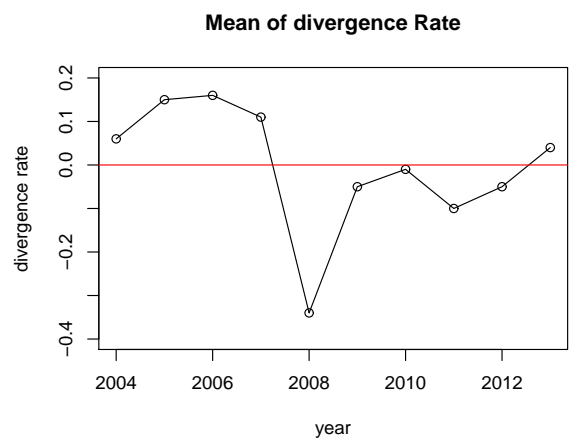}
\caption{The time series of mean of divergence rate in the period 2004 through 2013}
\end{center}
\end{figure}
\begin{figure}
\begin{center}
\includegraphics[width=8cm]{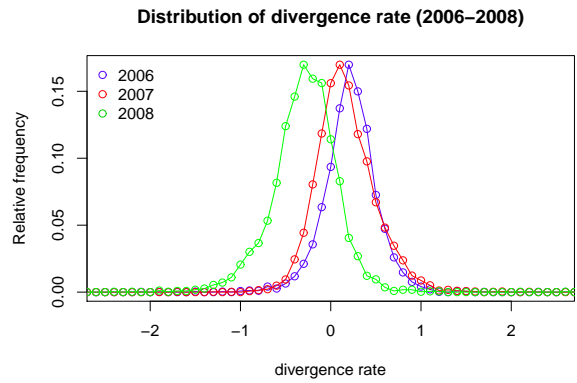}
\caption{The distributions of the divergence rate  for the period from 2006 through 2008. The distribution shifted negatively in 2008.}
\end{center}
\end{figure}
\begin{figure}
\begin{center}
\includegraphics[width=8cm]{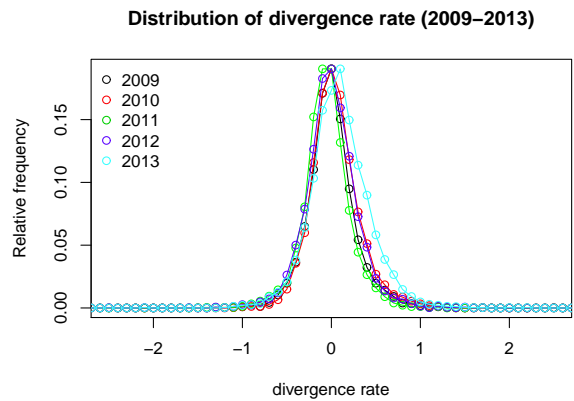}
\caption{The distribution of the divergence rate  for the period from 2009 through 2013. The distributions  were stable after 2008.}
\end{center}
\end{figure}

\end{document}